\documentstyle[epsf,floats,preprint,aps]{revtex}
\tighten

\def\beq{\begin{equation}}
\def\eeq{\end{equation}}
\def\grtsim
{~\lower0.6ex\hbox{\vbox{\offinterlineskip\hbox{$>$}\vskip1pt\hbox{$\sim$}}}~}

\def\lrtsim
{~\lower0.6ex\hbox{\vbox{\offinterlineskip\hbox{$<$}\vskip1pt\hbox{$\sim$}}}~}

\begin{document}
\begin{titlepage}

\begin{flushright}
 MIT-CTP-2501\\
 hep-ph/9512439
\end{flushright}



\LARGE
 \begin{center}
{\bf Supernatural Inflation: Inflation from Supersymmetry
with No (Very) Small Parameters
 }

\normalsize

 Lisa Randall \footnote{Work supported in part by the Department
of Energy under cooperative agreement DE-FC02-94ER40818, NSF
grant PHY89-04035, NSF Young Investigator Award, Alfred Sloan
Foundation Fellowship, and DOE Outstanding Junior Investigator
Award. } \\
Marin Solja\v{c}i\'{c} \footnote{Supported in part by MIT UROP}\\
Alan H. Guth \footnote{Work supported in part by the Department
of Energy under cooperative agreement DE-FC02-94ER40818.}\\
\vspace{.05in}
{\it Laboratory for Nuclear Science and Department of Physics \\
Massachusettts Institute of Technology \\
Cambridge, MA 02139}

\vspace{.15in}

\end{center}

\LARGE
\begin{center}
Abstract
\end{center}
\normalsize

Most models of inflation have small parameters, either to
guarantee sufficient inflation or the correct magnitude of the
density perturbations. In this paper we show that, in
supersymmetric theories with weak scale supersymmetry breaking,
one can construct viable inflationary models in which the
requisite parameters appear naturally in the form of the ratio of
mass scales that are already present in the theory.  Successful
inflationary models can be constructed from the flat-direction
fields of a renormalizable supersymmetric potential, and such
models can be realized even in the context of a simple GUT
extension of the MSSM.  We evade naive ``naturalness" arguments
by allowing for more than one field to be relevant to inflation,
as in ``hybrid inflation" models, and we argue that this is the
most natural possibility if inflaton fields are to be associated
with flat direction fields of a supersymmetric theory. Such
models predict a very low Hubble constant during inflation, of
order $10^3$-$10^4$ GeV, a scalar density perturbation index $n$
which is very close to or greater than unity, and negligible
tensor perturbations.  In addition, these models lead to a large
spike in the density perturbation spectrum at short wavelengths.

\vspace{0.25in}

\end{titlepage}

\section{Introduction}

Inflationary models \cite{in} in general require small parameters
in the particle theory Lagrangian, to provide the flat potential
needed for sufficient inflation and for the correct magnitude of
density fluctuations.  The need for unmotivated small parameters
tends to weaken the credibility of a theory, so one hopes that
the origin of these parameters can be understood.  It is
conceivable, of course, that the explanation lies beyond our
present understanding, just as we presently have no accepted
explanation of why the Yukawa coupling of the electron is $2
\times 10^{-6}$, or why the weak scale lies 17 orders of
magnitude below the Planck scale.  Nonetheless, it would be
encouraging to find that the small parameters required by
inflation could be obtained from small parameters that are
already essential to the particle theory, so that no additional
small parameters are introduced. Models in which the small
parameters arise as ratios of known particle physics mass scales
are particularly attractive \cite{ADFKRS}.

 {}From a  field theoretical perspective, it is difficult to see how
flat direction fields can be present in a nonsupersymmetric theory
(with the exception of Goldstone bosons, considered in Ref.
\cite{natural}) given the effect of radiative corrections .
We will assume therefore that the world is
supersymmetric, and ask whether an inflationary potential can arise naturally
in the context of the mass scales which we expect might be
present. Some natural candidates for these scales could be the
Planck scale $M_p \approx 10^{19}$ GeV, the GUT scale $M_{\rm G}
\approx 10^{16}$ GeV, the intermediate scale $M_I \approx
10^{11}$ GeV, and the supersymmetry breaking scale $m_{3/2}
\approx 1$ TeV.

However, straightforward considerations show that it is difficult
to implement this strategy.  The magnitude of density
perturbations points to the GUT scale as setting the energy
density during inflation, since $(M_{\rm GUT}/M_p)^2 \sim
10^{-6}$.
     Although suggestive, it is difficult to exploit this high scale
in an inflationary model.  In Ref. \cite{b1} it was argued that
the necessity for cancelling the cosmological constant after
inflation provides significant restrictions on a model in which
the inflation scale is greater than the supersymmetry breaking
scale.  Models have been suggested in which the energy density in
the early universe is well above the low energy supersymmetry
breaking scale, and might be governed for example by the value of  a moduli
field \cite{b2,thomas}. For these models, it must be checked that
the constraint of Ref. \cite{b1} is satisfied. Beyond this,
however, it is difficult to study such models in detail without a
concrete realization.

It might also be that the inflation scale is generated by GUT
physics.  An interesting example of this type of model is
Ref.~\cite{shafi}.  It is however questionable whether the GUT
scale exists as a fundamental scale of particle physics at all in
light of the doublet-triplet splitting problem \cite{r1}. 
Furthermore, it is likely that the inflaton field would be
charged under the GUT group and the reheat temperature would be
too high (in excess of the gravitino bound). 

In the context of supersymmetric models, an attractive scale for
the vacuum energy density during inflation would be set by the
intermediate scale, $M_I$.  This scale is very likely to be
present in a hidden sector model of supersymmetry breaking. It is
also the right energy scale for the potential associated with
moduli fields, which might be natural candidates for flat
directions.  The problem here, however, is that simple
dimensional analysis arguments (to be reviewed in Sec.~2) show
that density perturbations would generically be either far too
small \cite{rt,b1} or far too large, depending on assumptions.
For the case of inflation driven by a single moduli field, the
dimensional arguments show that the requirements of sufficient
inflation and correct density perturbations imply that 1) the
variation of the inflaton (moduli) field during inflation is of
order $M_p$, and 2) the energy density during inflation is of
order $M^4$ where $M\approx 10^{16}$ \hbox{GeV}. An energy
density of order $M_I^4$ would produce density perturbations too
small by about ten orders of magnitude.  For the case of a
chaotic inflationary scenario, the variation of the inflaton is
again of order $M_p$, but in this case the density perturbations
are much too large unless the quartic coupling $\lambda$ is about
$10^{-12}$.

In this paper, we show that the argument that inflation at an
intermediate scale is untenable lacks sufficient generality, and
can evaporate if one drops the assumption that inflation is
driven by a single scalar field.  We describe a class of
two-field models, for which dimensional analysis estimates show
that 1) the variation of the inflaton is of order $M_I$ or less,
and 2) the energy density during inflation is of order $M_I^4$.
We then go on to illustrate these ideas with models motivated by
supersymmetry with soft supersymmetry breaking. We will find that
these models not only solve the naturalness problem of obtaining
sufficiently many e-foldings of inflation, but also generate very nearly the
correct size of density perturbations based on the parameters of
supersymmetry breaking. We therefore refer to our models as
``supernatural" inflation.

This model contains a similar structure to the ``hybrid"
inflation models, proposed by Linde
 and studied by Copeland, Liddle, Lyth, Stewart, and Wands
\cite{hybrid}.  The fact that the standard dimensional
naturalness arguments for the number of $e$-foldings and for
$\delta \rho/\rho$ do not apply, and that the Hubble scale during
inflation will be low was also clearly recognized by these
authors. Our point here is to emphasize that the most natural
scales for successful implementation of two field inflation of
the ``waterfall" type are the scales associated with
supersymmetry breaking and the Planck scale.  Furthermore, our
models more accurately reflect masses and couplings associated
with flat direction fields, and we will motivate the parameters
and potential we use by consideration of flat directions in the
\hbox{MSSM}.  Hybrid inflation in the context of SUSY leads one
to the interesting conclusion that the Hubble scale during the
inflation which established the density perturbations might have
been of order $10^3$--$10^4$ GeV, rather than $10^{13}$ GeV.

In the following section, we present the general arguments for
why supersymmetry scales do not work in single field inflation
models. We then review the general idea of ``hybrid" or
``waterfall" \cite{hybrid} models, and show why the single-field
arguments do not apply to the two-field case.  In Section 3, we
present supernatural inflation models, in which we assume the
inflation sector consists of flat direction fields whose
potential is generated through supersymmetry breaking and
nonrenormalizable operators.  We derive the constraints on
parameters consistent with the requisite number of $e$-foldings
and density perturbations.  In Section 4 we explore the
possibility of a renormalizable coupling between the flat
direction fields.  In the following section we motivate the
models of Sections 3 and 4 by briefly considering flat direction
fields in the standard model, and present an illustration of the
model of Section 4 in the context of a GUT extension of the
\hbox{MSSM}.  In Section 6, we give details of the evolution of
the two fields in our models.  We analyze the density
perturbations that result from this evolution, and discover a
novel spike that is predicted to appear at short wavelengths.
Such a spike could lead to overproduction of black holes, but we
show in Section 7 that existing constraints on black holes are
satisfied for the parameters of interest.  In Section 8, we show
that the thermal production of gravitinos is also not a problem
in our models.  In the following section we show that successful
baryogenesis can be accomplished in the context of late
inflation.  In Section 9, we discuss miscellaneous aspects of
our models, and in Section 10, we conclude.

\section{One vs. Two Field Inflation}

We begin this section by reviewing the ``standard"
arguments  for why the inflaton in ``natural" inflationary
models varies on the scale $M_p$ and why
the scale for the energy density should be larger
than the intermediate scale in inflationary models
with a single field.

For the purposes of these dimensional arguments, we first assume
the potential takes the form
\beq
{\cal V}=M^4 {\cal G}(\phi/f)
\eeq
where ${\cal G}$ is a bounded function of order unity.  Here we
have in mind for example a moduli {field,} with $M\approx M_I$.
If we assume the slow roll equation of motion
$3H\dot{\phi}=-{\cal V'}$, where $H$ is the Hubble constant
during inflation, the number of $e$-foldings is
\beq\label{neqn}
N=\int H dt=\int d\phi {H \over \dot{\phi}}=- \int d\phi
{{\cal V} \over M_p^2 {\cal V'}}\approx-  {\Delta \phi \over
M_p} {f \over M_p} {{\cal G}\over {\cal G'}}
\eeq
There are essentially two possibilities.  If ${\cal G}$ is a
bounded function, and ${\cal G'}$ is not very tuned to have very
flat sections, one is in the regime of what might be expected for
a moduli type field. In this case, the requirement of about 60
$e$-foldings of inflation favors $f$ of order $M_p$ and a change
in $\phi$ during inflation at least of order $M_p$. Even when
this is satisfied, some tuning of the potential is required.

The alternative possibility is that one is in a chaotic
\cite{chaotic} inflationary scenario, in which case ${\cal G}$
will be dominated by monomial behavior for sufficiently large
field, and ${\cal V'}\approx {\cal V} /\phi$.  In this case, $f$
is not defined, but one would still conclude $\Delta \phi
\approx M_p$.

Density fluctuations are also readily estimated under the
assumed form of the potential.  They are given by
\beq
{\delta \rho \over \rho}\approx {H^2 \over \dot{\phi}}\approx
{H^3 \over {\cal V'}}\approx \left({M \over M_p}\right)^2 {f
\over M_p} {{\cal G}^{3/2} \over {\cal G'}}
\eeq
Assuming a potential of the moduli type, with ${\cal G}$ and
${\cal G'}$ of order unity and $f$ of order $M_p$, we find that
${\delta \rho \over \rho} \approx \left({M \over M_p}\right)^2$
favoring $M \approx 10^{-3} M_p$. Detailed calculations might
change $M$ by an order of magnitude or so, but it is clear that
$M \approx 10^{11}{\rm \ GeV}\approx M_I$ is strongly disfavored.

In a chaotic scenario on the other hand, one would conclude that
the density fluctuations are too large unless there is a small
parameter. For example, a simple dimensional argument would lead
to the conclusion that for $V=\lambda \phi^4$, $\lambda \approx
10^{-12}$.  Without further motivation for these small numbers,
such a potential seems unlikely.

So one is led to the conclusion that it is difficult to naturally
obtain sufficiently many e-foldings and the correct
magnitude of density perturbations, without invoking either
small numbers or a new mass scale.

It is apparent, however, that there is a loophole in the above
argument.  From Eq.~(\ref{neqn}) it is clear that the constraints
on $f$ and $\Delta \phi$ during inflation arise because it is the
same potential ${\cal V}(\phi)$ that controls the inflation rate
$H$ and the speed of the inflaton field $\dot \phi$.  These
constraints can be avoided, therefore, if the energy density
during inflation is provided from some source other than the
scalar field which rolls and controls the ending of inflation.

The simplest way to implement this idea would be with two fields.
This idea is essentially that first proposed by Linde
\cite{hybrid} as ``hybrid" inflation or ``waterfall" models.
There are two fields $\psi$ and $\phi$. The first field, which we
call the inflaton, has a very flat potential. It starts at a
large field value, and slowly rolls (via its classical field
equations) to the origin.

The second field, $\phi$, has a potential whose minimum is far
from the origin.  In most previous incarnations of hybrid
inflation, the scale of variation of this field is $M_I$, though
in our models the scale will be $M_p$.  When $\psi$ has large
field value, it gives a {\it positive} mass squared term in the
$\phi$ potential at the origin, so the classical field equations
{push} $\phi$ to the origin.  When $\psi$ gets sufficiently small
(of order $M_I$ or less in our models), the {mass squared} of
$\phi$ goes negative, and $\phi$ makes the transition from the
origin to $M_p$.

The key feature of this model is that the energy density during
inflation is dominated by the potential energy of the $\phi$
field at $\phi=0$.  There are no tunings in the $\psi$ potential
to get a small mass during inflation and a large mass afterwards,
since its mass is always small, as is its potential energy.
Because $H$ depends on the value of $\phi$ and is not determined
by the field $\psi$, which  acts as a switch to end inflation, the
naive estimates do not apply.

The second key feature of this model is that the ending of
inflation is controlled by when the $\phi$ mass squared at the
origin changes sign.  One can obtain a large number of
$e$-foldings with the variation of the inflaton field $\psi$ much
less than $M_p$.  Let us see this explicitly.

We  assume a potential which takes the form
\beq \label{pot}
{\cal V}=M^4 {\cal G}(|\phi|/f)+g(|\phi|, |\psi|)+m^2 |\psi|^2
\eeq
where $M$ and $f$ are to be determined, the function $g$
is the term responsible for the $\psi$ dependence of the $\phi$
mass, and $m$ is of order  $m_{3/2}$.

We now have
\beq
N \approx \int d\psi {H \over \dot{\psi}}\approx - \int {H^2
d\psi \over m^2 \psi}\approx {M^4 \over m^2 M_p^2}
\ln\left({\psi_{init} \over \psi_{final}}\right)
\eeq
Notice that the scale $M$ in the numerator is independent of the
mass and coupling of the $\psi$ field (in the limit that the
$\psi$ contribution to the energy density is small) so that the
previous arguments for one-field inflation no longer apply.
Clearly for inflation to give several $e$-foldings requires only
that $\psi$ changes by an order of magnitude, and that $M^4
\grtsim m^2 M_p^2$. No inflaton variation of order $M_p$ is
required, and so far, it seems $M\approx M_I$ could be a good
choice.

Let us now consider density fluctuations under the same
assumed form for the potential.
We find
\beq\label{6}
{\delta \rho \over \rho}\approx {H^2 \over \dot{\psi}}\approx
{H^3 \over m^2\psi}\approx{M^6 \over M_p^3 m^2 \psi}
\eeq
The point is that the numerator $H^3$ has its scale set by the
$\phi$ potential energy while the denominator is determined by
the $\psi$ field. We construct a model so that $\psi$ at the end
of inflation is of the order $M_I$ or smaller.  If we also take
$M \approx M_I$, we find $\delta \rho /\rho \approx M_I/M_p$ or
bigger (rather than $(M_I/M_p)^2$ as was the case in single field
models).  Although the coupling between the fields can have a
coefficient which varies by many orders of magnitude, as does
$\psi$ in eqn. (\ref{6}), the strong $M$ dependence of Eqn.
(\ref{6}) allows for agreement  with the COBE constraint with only a
relatively small $M$ variation. This is very promising from the
perspective of relating inflation models to real scales of
particle physics. To answer the questions of how well these ideas
really work, and how constrained the parameters of the models
really are, requires a detailed investigation of particular
examples of these ideas.

\section{ Supernatural Inflation }

We define Flat Direction Hybrid Inflation (FDHI) models as those
motivated by the properties of moduli fields or flat directions
of the standard model.  For moduli fields with no gauge charge or
superpotential, the whole potential arises from the Kahler
potential once supersymmetry is broken.  This potential for
$\phi$ and $\psi$ will take the form
\beq
V(\phi,\psi)=M_I^4 f(\phi/M_p,\psi/M_p)
\eeq
where the dimensionless coefficients in $f$ should be of order
unity.

However, it is clear that a model of this sort will not give rise
to inflation with sufficiently large density fluctuations, since
during the relevant period $\psi$ will typically be of order
$M_p$, and the resulting $\delta \rho / \rho$ will be of order
$(M_I/M_p)^2$.  We conclude that it is essential to have an
additional interaction between $\psi$ and $\phi$.  In this model,
we assume the existence of a superpotential which couples $\psi$
and $\phi$ but which is suppressed by a large mass scale $M'$.
For standard model flat directions, such higher dimension
operators are to be expected, with $M'$ equal to $M_p$, $M_G$, or
some dynamical scale.  In the case of moduli fields, it might be
that this scale is of dynamical origin; one can readily determine
how the answer changes with the form of the superpotential and
the size of the mass scale.

We therefore assume the presence of a superpotential.  The
example we take is
\beq
W={\phi^2 \psi^2 \over 2 M'}\label{supp}
\eeq

We now need to specify the form of the supersymmetry breaking
potential.  We assume both $\psi$ and $\phi$ have mass of order
the soft SUSY breaking scale of order 1 TeV (where we will need
to test the consistency of this assumption). We assume that the
potential for the $\psi$ field gives a positive mass squared at
the origin, while the $\phi$ field has negative mass squared at
the origin. Furthermore, we assume that the cosmological constant
is zero at the minimum of both $\psi$ and $\phi$. The
specific form of the potential we choose is
\beq
V=M^4\cos^2\left({|\Phi |\over f}\right)+m_\psi^2
|\Psi|^2+{|\Phi|^4|\Psi|^2+|\Psi|^4|\Phi|^2 \over M^{'2}}
\label{potential}
\eeq
where again we assume (and verify for consistent inflation)
$M\approx M_I$.  When the parameters are motivated by
supersymmetry breaking, we refer to our models by the name
supernatural inflation. We will see that one very naturally
obtains the correct magnitude of density perturbations, and
sufficiently many e-foldings of inflation, using parameters and a
potential which are well motivated in supersymmetric models.

For the purpose of the inflationary model, the
scalar field can be assumed to be real.  Defining $\Psi \equiv
(\psi + i \psi_I)/\sqrt{2}$ and a similar equation for $\Phi$,
the potential for the real fields becomes
\beq
V=M^4 \cos^2\left({\phi/\sqrt{2} f } \right)+{m_{\psi}^2\over 2} \psi^2
+{\psi^4\phi^2+\phi^4 \psi^2 \over 8 {M'}^2}
\eeq
The $\psi$ mass is $m_\psi$ and the magnitude of the
 {(imaginary)}
$\phi$ mass term (at the origin) is $m_{\phi}\equiv M^2/f$.
During inflation, $\phi$ is confined near the origin. The field
$\psi$ slowly rolls towards the origin and inflation ends about
when $\psi=\psi_c= \sqrt{2M'm_{\phi}}$.  It will turn out that
either $m_\psi/m_\phi$ or $M'/M_p$ is small, so that during
inflation the term $m_\psi^2 \psi^2$ is small relative to $M^4$.
The Hubble parameter during inflation is therefore approximately
$H=\sqrt{8 \pi/3}M^2/M_p$.

We expect $f$ is of order $M_P$, or equivalently, $m_\phi^2$ is
of order $m_{3/2}^2$.  Although it looks like we took a very
special form for the $\phi$-potential in Eq.~(\ref{potential}),
the use of the cosine is not essential.  As can be seen from a
Taylor expansion, only at the very late stages of inflation are
terms other than the constant and mass term relevant. We
could equally well have specified a potential which is truncated
at fifth order in the fields, or which has different higher order
terms. Although both $\psi$ and $\phi$ might be moduli or
standard model flat direction fields, we assume their potentials
are of very different form; the particular case we assume is
illustrative of how a model could work.

The  constraint from density perturbations in the slow-roll
regime is \cite{spectrum,cobe}
\beq
{V^{3/2}\over {\tilde M_p^3 (dV/ d\psi)}}=6 \cdot 10^{-4}
\eeq
where $\tilde M_p \equiv M_p/\sqrt{8\pi}$.
This gives the constraint
\beq \label{form}
{M^5 \over m_\psi^2 M_p^3}\sqrt{f \over M'} e^{rN}=6.7 \cdot 10^{-6}
\eeq
where
\beq
r=-{3 \over 2}+\sqrt{{9\over 4}+\mu_\psi^2}\approx {\mu_\psi^2\over 3}
\label{r}
\eeq
where the approximation in Eq.~(\ref{r}) is required if the slow
roll conditions are satisfied.  Here we have defined
$\mu_\psi=m_\psi/H$ and have measured time in $e$-foldings away
from the time $N=0$ when $\psi=\psi_c$ (where inflation ends at
positive $N$).  It is clear that a lower $M'$ makes the value of
$\psi$ at the end of inflation lower, which in turn {\it
increases} the density perturbations. The exponential in
Eq.~(\ref{form}) determines the scale dependence of the density
perturbations, characterized by the scalar index.

The scalar index $\alpha_s$ is readily determined from the scale
dependence of the density perturbations to be $-\mu_\psi^2/3$.
This can be seen directly from the formula for density
perturbations above. Alternatively, it is extracted from the
general formula \cite{turner}
\beq
n=1-2\alpha_s=1-3\left({V'\over V}\right)^2+2 {V'' \over V}
\eeq
where dimensionful factors should be compensated by
$\tilde M_p$. Notice that the second term is negligible for
all models for which the inflaton field value is much less than
$M_p$.  This is readily seen from
the fourth expression in Eqn. \ref{neqn},
which implies $\tilde M_p V'/V \approx \Delta \phi / (N M_p)$.
The third term is positive in our model, because the
inflaton field rolls toward, rather than away, from the origin
during the end of inflation.

We see for this model that $n$ is always greater than 1, and is
very close to 1 for small $\mu_\psi$, which is the case for large
$M'$.  This differs from the usual prediction for new inflation
or chaotic inflation models.  The current upper bound on $n$ is
uncertain as is summarized in Ref. \cite{cgl}.  These bounds,
along with the validity of slow roll, prevent too large values of
$\mu_\psi$. {}From Figures 1-3, we see large $\mu_\psi$ is
permitted only for the smallest value of $M'$, where the bound on
$n$ will provide an additional constraint.

The fact that $n$ is greater than or close to 1  is a  characteristic
feature of our models, which should help them to be
distinguishable in the future, when a good measurement
of the CMBR is obtained.

Another distinctive feature of these models is that
the  ratio of the tensor to scalar contribution to the quadrupole
\beq
{\cal R}={T \over S}\sim  \left({V'\over V}\right)^2 \approx 0.
\eeq
Again this follows from the small value of the inflaton field
$\psi$ near the end of inflation.

As we have argued in the first section, models of inflation
which have only a single field should have the inflaton
field taking a value of order $M_p$ near the end of inflation
if 50 e-foldings are to be obtained without fine tuning.
The combination of negligible ${\cal R}$ and $n$ never
below 1 are distinctive features of these models
which should help distinguish them from
other possible inflationary models in the future.

\begin{figure}
\centerline{\epsfbox{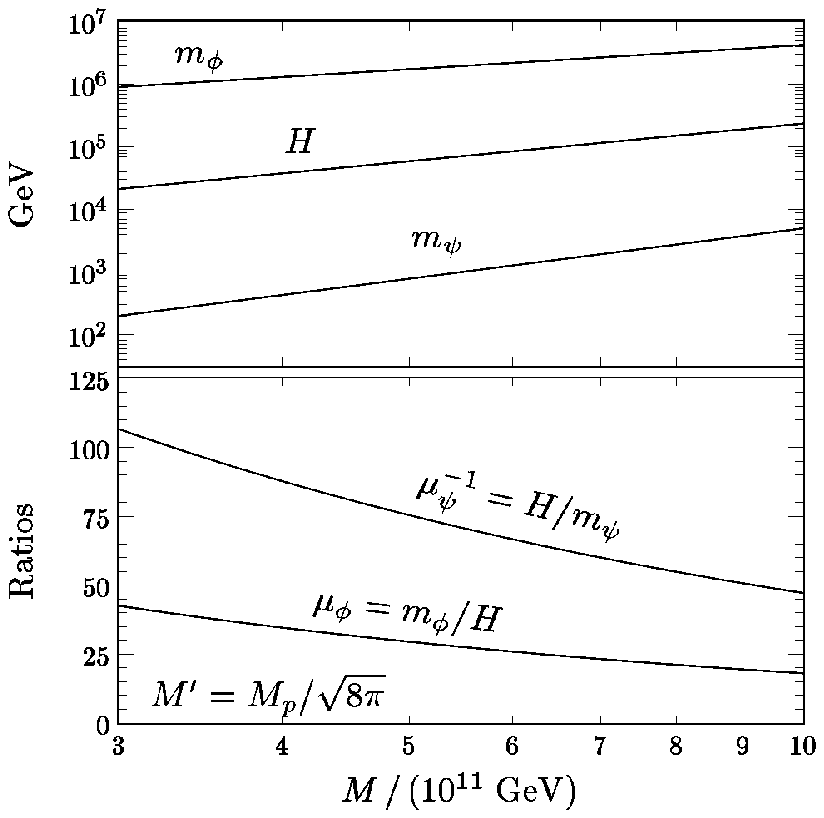}}
\caption{Parameter choices for supernatural inflation with $M'$
at the Planck scale.  Here $\mu_\psi$ is chosen
to give the correct magnitude of density fluctuations for
the minimum $\mu_\psi$ consistent with a sufficiently
rapid end to inflaton.}
 \vskip 0.2truein
\centerline{\epsfbox{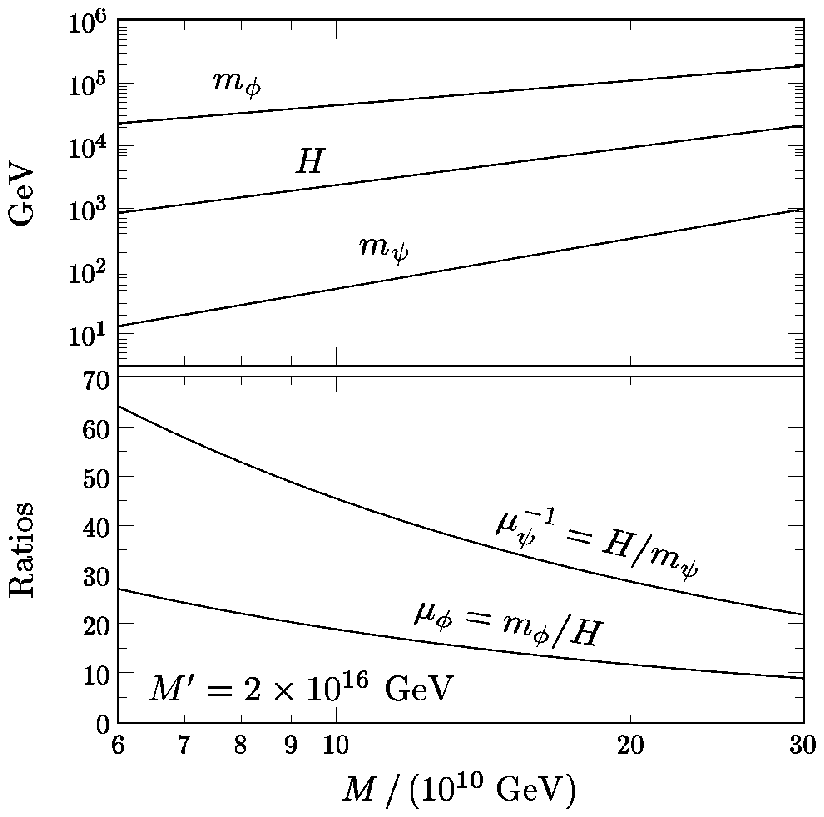}}
\caption{Parameters for supernatural inflation with $M'$ at the
GUT scale.  The values were chosen by the same criteria used in
Fig.~1.}
\end{figure}

\begin{figure}
\centerline{\epsfbox{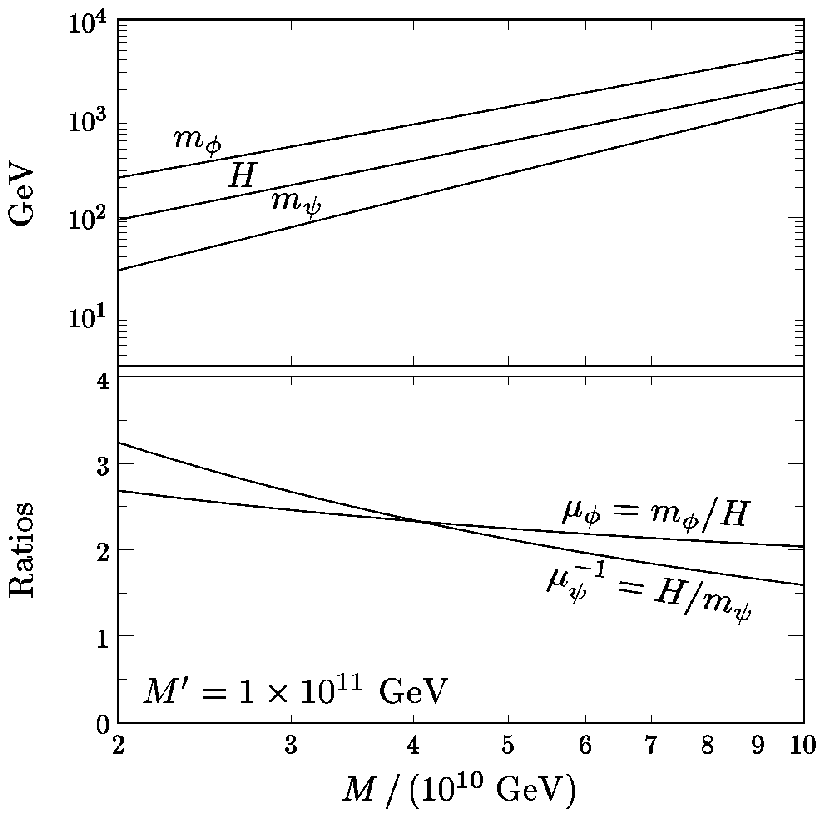}}
\caption{Parameters for supernatural inflation with $M'$ at the
intermediate scale.}
\end{figure}

In Figs. 1--3, we show   values of the parameters when $M'
\approx M_p$, $M'\approx M_{GUT}$ and $M'\approx M_I$
respectively.  The values shown were found by imposing the
correct magnitude of density fluctuations and choosing the
minimum $\mu_\phi \equiv m_\phi/H$ consistent with a sufficiently
rapid end of inflation (see Sec.~6).  We chose the range of
$M$ to optimize parameters.  Smaller $M$ would increase the
values of $1/\mu_\psi$ and $\mu_\phi$.  Large $M$ would improve
(that is, decrease) these ratios but would make the masses
uncomfortably large relative to the TeV scale.  We find that
smaller $M'$ gives more natural ratios for the mass to the Hubble
scale, though in all cases a ratio of less than 100 can be
obtained.

These constraints assumed that the contribution of $\phi$ to
density perturbations was small. In order to check the
consistency of this assumption, we need to consider the evolution
of $\psi$ and $\phi$ in the late stages of inflation.  It will
turn out that inflation must end reasonably quickly after $\psi$
reaches $\psi_c$ so that perturbations exit the horizon while the
$\phi$ field is still confined to the origin.  This gives a lower
bound on $\mu_\phi$. In Sec.~6, we will investigate the $\phi$
mass constraint in detail.

\section{Another Model}

In Section 3, we investigated the possibility that there is a
nonrenormalizable superpotential.  However, it is frequently the
case that flat directions lift each other; that is, the
renormalizable potential does not permit certain field directions
to be simultaneously flat. In this section, we present an
alternative model with a renormalizable potential. It will turn
out that this model requires a small coupling in the potential.
We will motivate this assumption in Section 5, where we consider
particular choices of $\psi$ and $\phi$ chosen from the
supersymmetric standard model where we will show that the small
coupling can actually be related to one of the known small Yukawa
couplings! Once we assume this small parameter (again an
unexplained but perhaps necessary parameter of the MSSM) we will
find that $\mu_\psi$ and $\mu_\phi$ can both be close to unity.

So we take the potential to contain the soft supersymmetry
breaking terms as before but to contain a renormalizable coupling
between $\psi$ and $\phi$. Specifically
\beq
V=M^4 \cos^2\left({\phi/\sqrt{2} f } \right)+{m_{\psi}^2\psi^2
\over 2} + \lambda^2 {\psi^2 \phi^2 \over 4}
\eeq

This model has the essential features of the FDHI model of the
previous section. The difference is the value of $\psi_c$ which
in this model is
\beq
\psi_c={\sqrt{2} m_\phi \over \lambda}
\eeq
The density fluctuations give the constraint
\beq
{\lambda H^3 e^{rN} \over m_\psi^2 m_\phi}=1.6 \cdot 10^{-4}
\label{eq:18}
\eeq
where
\beq
r=-{3 \over 2}+\sqrt{{9\over 4}+\mu_\psi^2}\approx {\mu_\psi^2\over 3}
\eeq

\begin{figure}
\centerline{\epsfbox{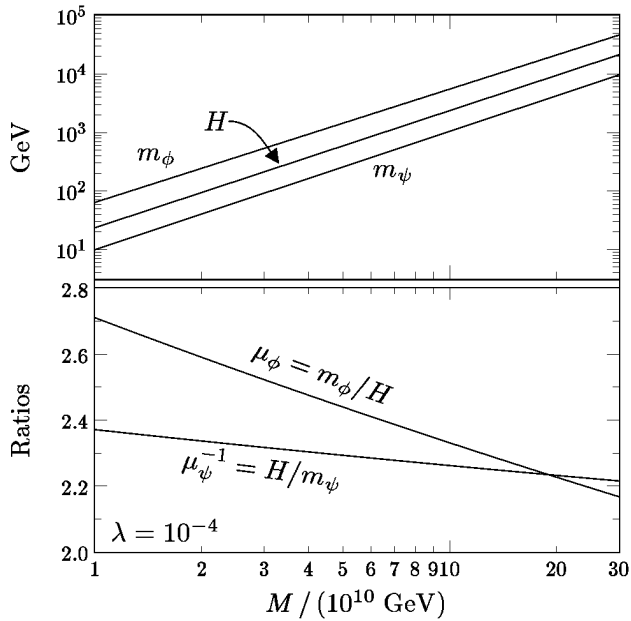}}
\caption{Parameter choices for supernatural inflation with
renormalizable couplings, for $\lambda=10^{-4}$.  Like the
previous figures, this graph shows the parameters associated with
the minimal allowed value of $\mu_\phi$.}
\end{figure}

If $\lambda$ is of order unity, to satisfy Eq.~(\ref{eq:18})
requires $\mu_\phi>10^3$. However, if $\lambda\approx
10^{-4}-10^{-5}$, the model works perfectly with $\mu_\psi$ and
$\mu_\phi$ both of order unity. In Fig.~4 we plot the parameters
of the model for $\lambda= 10^{-4}$, where again we have chosen
the minimum $\mu_\phi$ consistent with a sufficiently quick end
to inflation.  We see there is virtually no fine tuning, so long
as a small $\lambda$ exists(!). In the following section, we
explain why such a small value for $\lambda$ is not necessarily
unexpected.

\section {Examples of ``Flat Direction" Interaction Potentials}

Up to this point we have considered abstractly how to make a
successful inflation model premised on properties motivated by
flat direction fields. In this section, we motivate the sort of
models we have considered by demonstrating examples of flat
direction fields in the MSSM whose couplings are those required
for a successful supernatural inflation model.  However, because
the $\phi$ field minimum is not at zero, which in the context of
the standard model would imply large gauge symmetry breaking, the
minimal standard model is not appropriate.  Nevertheless, in
simple extensions of the standard model, there are neutral fields
which can play the role of the $\phi$ field. We will present an
example of a GUT extension of the MSSM which could contain
appropriate ``$\psi$'' and ``$\phi$'' fields.

We first present a few interesting cases of the form of the
superpotential which arises in the context of the MSSM. Other
flat direction possibilities could motivate further
generalizations of the models we have studied.

Flat directions of the MSSM were considered in Ref. \cite{drt}
and a complete analysis was given in Ref. \cite{gh}. They
can be parameterized by gauge invariant combinations
of fields. For example, the flat direction $H_u^1=H_d^2=\phi$ will be
denoted as $\phi=H_u^1H_d^2$ where we have given explicit
gauge indices.  In each case, it is necessary to explicitly
check for $F$-flatness by examining the form of the superpotential.

Example: $\psi=\bar{u_1}^1\bar{d}_1^2\bar{d}_2^3$, $\phi=Q^1_2
\bar{d}_3^1 L_2$ $W={Q\bar{u} Q \bar{d}/M'}$.  The superscripts
here refer to the color indices and the subscripts to flavor
indices. We haven't specified the indices in the superpotential
where in principle all allowed contractions can appear in
different terms.  This model would agree with the physics of our
first model of inflation, except that the superpotential actually
takes the form $\psi^2 \phi X/M'+\alpha\phi^2 \psi X/M'$, where
$X$ is a field which is not flat and which stays at zero during
the relevant stages of inflation.  This can be seen by explicitly
substituting $\bar{u}_1^1=\bar{d}_1^2 =\bar{d}_2^3=\psi$ and
$Q^1_2=\bar{d}_3^1=L_2=\phi$ into all possible terms of the above
form in the superpotential (that is with arbitrary flavor and
allowed gauge contractions). The potential will be as we have
studied, except that the coefficients of the $\psi^4 \phi^2$ and
$\phi^4 \psi^2$ terms need not be identical, since they arise
from distinct terms in the superpotential, and there is a cross
term which changes the exact evolution of $\phi$ but in no
significant way, which arises because the separate superpotential
terms can depend on the same $X$.

Example: $\psi=LH_u$, $\phi=H_u H_d$

In this example, SU(2) $D$ terms will lift the flat direction.
This can be seen by solving for the fields in terms
of the flat directions and substituting into the $D$
terms of SU(2). It can be seen that the $D$ term
does not vanish, but can involve a $\psi^2 \phi^2$
cross term, suppressed only by $g_2^2$, where $g_2$
is the SU(2) gauge coupling.
As discussed in Section  4, a model of this sort with a large
gauge coupling can work, but requires tuning $\mu_\phi$.

Example: $\psi=\bar{u} \bar{d}\bar{d}$, $\phi =LH_u$,
$W=\lambda_u Q_u H_u \bar{u}$

This example realizes perfectly our scenario with a
renormalizable potential generated by a small Yukawa coupling,
but not a large gauge coupling.  In fact, if it is indeed the up
quark in the $\psi$ field, the potential works as well as could
be hoped, since it depends on $\lambda_u^2 \phi^2 \psi^2$ which
as we have shown gives $\mu_\psi$ and $\mu_\phi$ of order unity.
In this model, the small size of density fluctuations arises as a
natural consequence of the small up quark Yukawa coupling!

There are in fact other interactions in this model, with the
field $Q_u$. However $Q_u$ is not a flat direction and is assumed
to be zero (or small) throughout inflation, so that it is
irrelevant to the analysis.

In fact there are many examples of the above type. Even
with somewhat bigger Yukawa coupling, the correct
magnitude of density fluctuations can be obtained at the
expense of a larger ratio of $\mu_\phi$.  This is
probably the nicest possibility for realizing the inflationary
scenario we have outlined, because the only small
numbers are those already present in the form of Yukawa couplings.
There are no unlikely assumptions required for the
correct magnitude of density perturbations and a sufficiently
rapid exit to inflation.

The problem with the MSSM  as the source
of inflaton candidates, as we have already stated,
  is that $\phi$ has a nonzero expectation value
at the end of inflation.  Because $\phi$
in general carries standard model
gauge charge, this
is not permitted. However, in GUT or other
generalizations of the MSSM (or in models with completely independent
fields which do not carry standard model gauge charges)
one can readily realize the scenario we outlined.

{ Model:} Consider a generalization of the standard SU(5)
GUT theory to SU(6), where now the  Higgs fields are
in the $6$, $\bar{6}$, and $35$  ($H$, $\bar{H}$,
$\Sigma$) representations of SU(6). The matter consists of
 three generations of $15+\bar{6}+{\bar{6}}'$.
This model has been well studied in the context of solving
the doublet triplet splitting problem\cite{pgb}. Examples
of specific models with the requisite accidental
symmetry were presented in Ref. \cite{cr}.

Here we assume there is no superpotential for the $H$ and $\bar{H}$ fields,
but that the potential created by the soft supersymmetry
breaking terms is minimized at $\langle{H} \rangle\sim \langle{\bar H}\rangle
\sim M_p$.  Notice that this will break SU(6) to SU(5) which
can survive to the GUT unification scale, and is therefore
phenomenologically consistent. The $\Sigma$ field acquires
a vacuum expectation value of order $M_G$. We assume
that the $\Sigma$ field acquires this expectation value
through renormalizable interactions, and is therefore not flat,
and furthermore has reached its true minimum at the time of inflation.

Now consider $\phi=H^0\bar{H}^0$ and $\psi=15_2^{12} \bar{6}_1^2
\bar{6'}_3^1$ where we have labeled the matter according to its
generation number (subscript) and according to the SU(6) index
(superscript) (where 0 indicates the SU(5) neutral direction).

The superpotential which is required to complete the model can
readily be chosen in accordance with the requirement of a small
Yukawa coupling. To explicitly write the term is subtle however
for two reasons. First, the leading (renormalizable terms) which
give mass are the terms which make the $\bar{6}'$ fields heavy
and the term which gives the top quark a mass. However to give a
renormalizable top quark coupling requires that the top be in a
20 of SU(6). All other masses arise from nonrenormalizable
operators, and therefore appear more complicated. However,
because of the large expectation values of the $\Sigma$, $H$, and
$\bar{H}$ fields, these terms reduce to ordinary Yukawa
couplings.

A toy model which would give the necessary Yukawa coupling would
be $W=\lambda_\mu 15 _2\bar{H} \bar{6}_2$, where
$\lambda_\mu\approx 10^{-3}$.  This example resembles the up
quark example. However, this is only a toy model because such a
term actually gives the $\bar{6}'$ a mass, and in fact defines
the $\bar{6}'$ fields.  If this Yukawa coupling happens to be
small, the density fluctuations would be small. Since we know
little about the extra quark Yukawa coupling, we give a model
involving the known quark mass parameters.

The higher dimension term $W=V_{bc} (20 \Sigma) ({H \over M_P})
15_2$ can generate the mixing angle between the second and third
generation. The effective Yukawa coupling between the flat
direction fields from this term is $V_{bc} {\Sigma/M_p}$ which is
about $10^{-4}$ (since the $\Sigma$ VEV breaks SU(5) to ${\rm
SU}(3) \times {\rm SU}(2)\times {\rm U}(1)$ at the GUT scale).
The density fluctuations in this model are then naturally of
order $10^{-4}$.

The last model works very well, as is illustrated in Figure 4. It
is extremely interesting that the inflation scenario we have
devised can be explicitly realized in the context of a known
model of particle physics.

\section{$\psi$ and $\phi$ Evolution}

In this section we will discuss the details of the evolution of
$\psi$ and $\phi$. We can then derive the constraint on the
$\phi$ mass.

As we have argued, inflation ends at about the time the $\phi$
squared mass changes sign at the origin so that $\phi$ will roll
towards its true minimum. However, because the $\phi$ mass is not
large compared to $H$ as in previous implementations of the
hybrid inflation scenario \cite{hybrid}, a careful study is
required to ensure that inflation ends sufficiently rapidly that
our formula for density perturbations applies. We will see that
if inflation ends too slowly, perturbations will leave the Hubble
radius when the $\phi$ fluctuations are large.  In this scenario,
if these fluctuations were on measurable scales (say greater
than 1 Mpc and smaller than $10^4$ Mpc) either the size of
density fluctuations or the deviation from a scale invariant
spectrum would exceed the experimental bound.  By a detailed
study of the combined evolution of the $\psi$ and $\phi$ fields,
we determine the necessary constraint on the $\phi$ mass for
consistency of our model. However, throughout this section, it
should be remembered that this constraint is only very important
when $\mu_\psi$ is small, because it slows the transition which
causes the end of inflation. When $\mu_\psi$ is close to unity,
values of $\mu_\phi$ near unity are also adequate for a rapid end
to inflation.  For this reason, we will focus in the discussion
here on the case of small $\mu_\psi$. We present in detail the
analysis for the model of Section 3, where the constraint is more
severe.  A similar analysis was done for the model of Section 4
in order to obtain Figure 4.

We first consider the time evolution of $\psi$. The equation of
motion for $\psi$ is
\beq
\ddot{\psi}+3H\dot{\psi}+m_\psi^2 \psi+{\phi^4 \over 4M'^2} \psi=0
\eeq
where the dot denotes derivative with respect to time $t$.

There are three relevant stages of evolution of the $\psi$ field.
In the early stage of its evolution when the $\phi$ field is
small (and so is $\phi^4/{M'}^2$), the $\psi$ field obeys the
slow roll equation of motion as it evolves towards the origin.
\beq
\psi(t)=\psi_c e^{-\mu_\psi^2 N/3 }
\label{psiexp}
\eeq
where $\psi_c=\sqrt{2M'm_\phi}$ is the value of $\psi$ when
$m_\phi^2(N=0)=0$ (at the origin) where we measure time in units
of $H^{-1}$. Eventually, the $\phi$ field will grow to a
sufficiently large value $\phi_c$, where the $\psi$ mass becomes
large, and the $\psi$ field acts as a coherent state of
oscillating particles with mass $m_\psi(t) =
\phi^2/2M'$.  (Here we use the argument $t$ to distinguish the
time-dependent physical mass of $\psi$ particles from the
time-independent mass parameter in the potential.) We define
$\phi_c $ by $\phi_c^2/2M' = H$.

In our numerical simulations, we replace the time evolution of
$\psi$ by the time evolution of its envelope at a time
sufficiently late that we can neglect the term $\psi^4
\phi^2/{8M'}^2$ in the potential.  The envelope $\psi_e$ obeys
the approximate equation of motion
\beq
\dot{\psi_e}=-\left({3 H \over 2}+{1 \over  2} \Gamma(t) \right)\psi_e
\eeq
where $\Gamma (t)$ is the $\psi$ decay rate, where the
time dependence arises from the time dependent mass. Without further
knowledge of the identity of the $\psi$ field the decay rate is
an unknown parameter of the theory. We constrain the model under
two reasonable scenarios for the $\psi$ decay. If $\psi$ has
renormalizable couplings to other fields, its decay rate can be
as large as $\Gamma_b\approx m_\psi(t)$. Of course there are
unknown coefficients to this estimate but this probably
represents the maximum possible rate. The true rate should lie
between $\Gamma_b$ and $\Gamma_l$, where
$\Gamma_l=m_\psi^3(t)/(M_p/\sqrt{8\pi}) ^2$. This latter decay
rate assumes no renormalizable couplings, but Planck suppressed
interactions which allow the $\psi$ field to decay. We evaluate
the final stages of evolution allowing for these two
possibilities for the decay rate.  After the time at which
$\Gamma_\psi\approx H$, the amplitude of the $\psi$ envelope is
quickly reduced to zero, and it is only the $\phi$ field which
remains.  Notice that $m(t)$ could exceed $M_p$
before the end of inflation.
  However, the $\psi$ field decays well
before this inconsistency in the expansion is reached.

Because of the strong dependence on the $\psi$ field of
$m_\phi(t)^2=m_\phi^2(1-\psi^4/\psi^4_c)$, the $\psi$ evolution
is critical to determining the $\phi$ evolution.  In the early
stage of the $\phi$ evolution, it can be described by a Fokker
Planck probability distribution $P(\phi,t)$. This distribution
will be centered at the origin, but the spread will determine the
effective amplitude of the field $\phi\equiv\sqrt{\langle\phi^2\rangle}$.
Eventually the effective amplitude will be sufficiently large
that the classical equations of motion take over. By considering
the exact solution to the Fokker Planck equation, we determine
the correct initial condition for the subsequent evolution of the
classical field equations.

The Fokker Planck equation in a de Sitter background with time
independent Hubble constant $H$ is \cite{fp}
\beq
{dP(\phi,t)\over dt}={H^3 \over 8\pi^2}\left[{d^2 \over
d\phi^2}\right]P(\phi,t) -{d \over d\phi}\left[{m^2(t)\phi\over
3H} P(\phi,t)\right]
\eeq
where we have assumed the slow roll equation of motion to be
valid. With the evolution of $\psi$ described by
Eq.~(\ref{psiexp}), the time-dependent mass of $\phi$ is given by
\beq \label{MassSquared}
m_\phi^2(t) = m_\phi^2 \left(1 - e^{-4\mu_\psi^2 H t/3 }\right)
\eeq

The remarkable thing is that this is solvable by a Gaussian, even
when the mass is time dependent.
\beq
P(\phi,t)={1 \over \sqrt{2\pi} \sigma(t)}e^{-\phi^2/2\sigma(t)^2}
\eeq
Here $\sigma(t)$ obeys the equation
\beq \label{x}
{d \sigma(t) \over dt}={H^3 \over 8\pi^2 \sigma(t)}+{m_\phi^2 (t)
\over 3H}\sigma(t)
\eeq
Define $S(t)=\sigma^2(t)$. Then
\beq \label{26}
{d S(t) \over dt}={H^3 \over 4 \pi^2}+ {2 m_\phi^2(t) \over 3H}
S(t)
\eeq
Notice that this equation is readily interpreted as the field
$\phi$ subject to the force from the classical potential (the
second term) in addition to the force driving Brownian motion due
to de Sitter fluctuations \cite{desitter} (the first term). This
equation is readily solved by finding the appropriate integrating
factor and imposing the boundary condition that at $t=-\infty$,
$\sigma(-\infty)=0$. The solution is
\beq \label{27}
S(N)={H^2 \over 4 \pi^2}\int_{-\infty}^N \exp\left\{ {1 \over 2
\xi^2}
\left[
e^{-4 \mu_\psi^2 N / 3 } - e^{-4 \mu_\psi^2 N' / 3 }
     + {4 \mu_\psi^2 \over 3} (N - N')
\right]\right\} dN'
\eeq
where $\xi \equiv \mu_\psi/\mu_\phi$ and $w \equiv 1/ \mu_\psi
\mu_\phi$.  The integrand has a peak at $N'=0$, with a width of
order $w$.  When $\xi$ is small, which is generally the case in
our models, the peak is nearly Gaussian and a saddle point
approximation becomes applicable.  So for $N$ somewhat bigger
than $w$ (so that the peak is covered by the integration) and
$\xi
\ll 1$, $S(N)$ is well-approximated by
\beq \label{ApproxS}
S(N)={3 H^2 w \over 4 \pi^{3/2} } \exp\left\{ {1 \over 2 \xi^2}
\left( e^{- 4 \mu_\psi^2 N / 3}
     + {4 \mu_\psi^2 N \over 3 } - 1 \right)
\right\} \approx {3 H^2 w \over 4 \pi^{3/2} }e^{4 N^2 /9 w^2}
\eeq
where the final approximation is valid if $w \lrtsim N \ll
1/\mu_\psi^2$.

It should be borne in mind however that the Fokker Planck
equation we used incorporated the slow roll equation of motion,
which is valid for $N$ small compared to $w^2$ Because it will
turn out $w$ is not large, the Fokker Planck description will be
valid only at early times.  We therefore use the Fokker Planck
equation to establish initial conditions, and then use the
classical field equations to describe the $\phi$ evolution which
we evolve numerically.

While Eq.~(\ref{27}) provides an analytic solution to the
differential equation (\ref{26}), the qualitative behavior of the
solution can be seen by looking at the differential equation
itself. For large negative values of $t$, $m_\phi^2(t)$ is large
and negative, providing a strong restoring force.  This period is
characterized by quasi-equilibrium evolution, in which the
restoring force holds $S(t)$ very close to its equilibrium value,
$- 3 H^4 / 8 \pi^2 m_\phi^2 (t)$, for which $d S/ d t$ would
vanish.  The spread of this equilibrium probability distribution
approaches zero in the asymptotic past, and grows monotonically
with time.  As $t$ approaches 0, however, this equibrium value of
$S$ diverges, and the quasi-equilibrium regime ends because
$S(t)$ is not able to keep up.  We can estimate when the
quasi-equilibrium regime ends by asking when the velocity of the
equilibrium value exceeds the diffusive velocity, $H^3/4
\pi^2$.  For small $\xi$, this happens when $N \equiv N_i = -w
\sqrt{9/8}$.  Then $S(t)$ starts to grow diffusively, increasing
linearly in time according to the first term on the right-hand
side of Eq.~(\ref{26}).  Neglecting the growth before $N=N_i$,
which, in practice, changes the result by a number of order
unity, we estimate $S(0)$ as $(H^2/4 \pi^2) (N - N_i) =
\sqrt{9/8}\,H^2 w/ 4 \pi^2$, which in the limit of small $\xi$
gives an answer a factor of $\sqrt{2 \pi}$ smaller than the exact
solution.

The diffusive regime ends when the second term on the right-hand
side of Eq.~(\ref{26}) becomes larger than the diffusive term.
This final phase can be called the classical regime, since the
second term represents purely classical evolution.  If only this
term were included, the Fokker Planck equation would describe an
ensemble of classical trajectories.  This classical behavior is
essential to our treatment of the problem, since it allows the
description at late times to join smoothly to the full classical
equations of motion which remain valid outside the slow-roll
regime.  The transition from the diffusive to the classical
regime can be estimated by the ``velocity matching criterion",
which is precisely when the two terms on the right-hand side of
Eq.~(\ref{26}) are equal, approximating the solution until this
transition by the diffusive relation $S(t)
\approx H^3 t / 4 \pi^2$.  In the limit of small $\xi$, this
velocity-matching condition holds at $N \equiv N_0 = w
\sqrt{9/8}$, and the value of the spread is given by $S(N_0) \equiv
\phi_i^2 = 3 \sqrt{2} H^2 w / 16 \pi^2$.  The classical regime
can be approximated by constructing a solution to the classical
equations for $\phi(t)$, starting from the initial condition
$\phi(N_0) = \phi_i$.
  If the asymptotic
behavior of this classical solution (in slow-roll approximation)
is compared with the asymptotic behavior of $\sqrt{S(t)}$ as
given by Eq.~(\ref{ApproxS}), it is found to be smaller by a
factor of $( 8 \pi e)^{1/4} \approx 2.9$.  In practice, we use the
Fokker-Planck equation to establish the
initial condition at $N_0$.In our numerical
calculations we corrected for this discrepancy by using the
initial condition $\phi(N_0) = \bar \phi_i = ( 8 \pi e)^{1/4}
\phi_i$.

We have determined the time evolution of $\phi$ and $\psi$
subsequent to the velocity matching time numerically. However, as
for $\psi$, the classical evolution of $\phi$ can be determined
very well analytically.  Again, we have to divide the analysis
into three stages, according to the behavior of $\psi$.

At early times, the equation of motion for $\phi$ is
approximately given by
\beq
{d \phi \over dN}={4 N\mu_\phi^2 \mu_\psi^2 \over 9}\phi
\eeq
which is solved by
\beq
\phi(N) =\bar \phi_i e^{2(N-N_0)^2/9w^2}
\eeq
where we have imposed the boundary condition $\phi=\bar \phi_i $
at $N=N_0$. This solution has assumed slow-roll which is only
approximately valid. This stage of evolution of $\phi$ lasts for
$N_1\approx w\sqrt{9/2}\sqrt{\log(\phi_c/
\bar\phi_i) }$
$e$-folds, where $\phi_c=\sqrt{2M'H}$. Numerically, we have found
this answer to be off by 1-3 $e$-folds due to the correction to
slow roll.

At later time, as discussed above, $\psi$ begins to oscillate.
Depending on the $\psi$ decay rate, there can be several
$e$-folds between this time and the time at which the $\psi$
field decays.  During this range of time, it can be checked that
the $\psi^4 \phi^2$ term is no longer important to the equation
of motion and that $\phi$ essentially keeps up with the minimum
of the potential
\beq
V=-{m_\phi^2 \phi^2 \over 2}+{\phi^4 \psi^2 \over 8 {M'}^2}
\eeq
and is
\beq
\phi(N)={\sqrt{2} m_\phi M' \over \psi(N)}={\sqrt{2} m_\phi M' \over
\psi_c}e^{3N/2}=\sqrt{m_\phi M'}e^{3N/2}
\eeq
which is valid when $\Gamma\ll H$. Finally, the $\psi$ field
decays. This occurs when $\Gamma\approx H$. For $\Gamma_b$, the
number of $e$-folds during this stage is approximately $N_2^b=0$.
For $\Gamma_l$, the value of the $\phi$ field when $\psi$ decays
is approximately $\phi_l=(H{M'}^3M_p^2/\pi)^{1/6}$.  The total
number of $e$-folds in this stage is approximately $2/3
\log(\phi_l/\phi_c)$ which is $N_2^l=(1/9)\log(M_p^2/8H^2\pi)$.

After $\psi$ decays, $\phi$ follows the $\psi=0$ equation of
motion according to
\beq
\phi(t)\propto e^{rN}
\eeq
where
\beq
r=\sqrt{\mu_\phi^2+{9\over 4}}-{3 \over 2}
\eeq
The number of $e$-folds in this stage is
$N_3=(1/r)\log(\phi_f/\{\phi_c,\phi_l\})$ depending on the decay
rate, where $\phi_f=\pi/\sqrt{2} f$.

In fact we have checked that the solution above gives the number
of $e$-folds for inflation to end correct to within 1-3
$e$-folds.

The reason we require an accurate determination of the number of
$e$-foldings required for inflation to end is that it must be
that the density fluctuations relevant for the observed physical
scales have exited the horizon while the $\phi$ field is in the
early quasi-equilibrium stage.  As we will now show, the quantum
fluctuations of the $\phi$ field during the diffusive regime
generate a spike in the density perturbation.  For the model to
be viable, it is important that this spike occurs at short
wavelengths, so as to avoid conflict with observations.

One interpretation of the source of density fluctuations is that
the end of inflation occurs at different times at different
points in space. The time delay function $\Delta \tau({\bf x})$,
multiplied by $H$, is of the order of $\delta \rho/\rho$ at the
time the wavelength re-enters the Hubble length.  For example, if
$\psi\sim e^{-m_\psi^2 N/3H}$ and $\Delta \psi \sim H/2\pi$, the
density fluctuation constraint would be
\beq
H \Delta \tau=H{\Delta \psi \over \dot{\psi}}={1 \over 2 \pi
\sqrt{3}} {V^{3/2} \over (M_p/\sqrt{8\pi})^3 V'}={6 \times
10^{-4}\over 2\pi \sqrt{3}}.
\eeq

Since $\phi$ has a more complicated evolution, the calculation of
density fluctuations at early times is subtle.  The problem is to
estimate the density fluctuations caused by quantum fluctuations
in $\phi$ \cite{desitter}, whose value at times near $t=0$ is
determined by the Fokker-Planck diffusion equation. Let us first
consider the density fluctuations in early stages of the $\phi$
evolution. Suppose at a given time $t_1$ the solution $P(\phi,t)$
is modified by displacing the entire probability distribution by
an amount $\Delta \phi$, so that $P_{new}(\phi,
t_1)=P_{old}(\phi-\Delta \phi,t_1)$. We now ask how this will
affect the time at which inflation ends. Because the $\psi$ and
$\phi$ equations of motion are effectively decoupled, we can
treat the $\psi$ field as uninfluenced, and treat the $\phi$
field as a free field evolving with time-dependent squared mass
given by Eq.~(\ref{MassSquared}).

We guess the solution is a shifted Gaussian,
\beq
P(\phi,t)={1 \over
\sqrt{2\pi}\sigma(t)}e^{-(\phi-\bar{\phi})^2/2\sigma(t)^2}
\eeq
We find that the Fokker-Planck equation is satisfied provided
that $\sigma^2(t)$ obeys Eq.~(\ref{26}), and $\bar \phi(t)$ obeys
the equation
\beq
{d \bar{\phi}(t) \over dt}={m_\phi^2(t) \over 3H}\bar{\phi}
\eeq
Imposing the initial condition $\bar \phi(N_1) = \Delta\phi$, the
solution to this equation is
\beq
\bar{\phi}(N)=\Delta\phi \exp\left[{1 \over 4 \xi^2}
\left(e^{-4\mu_\psi^2N/3}-e^{-4\mu_\psi^2N_1/3}+{4\over
3}\mu_\psi^2 \left(N-N_1\right)\right)\right]
\eeq
Since the entire distribution is shifted uniformly by
$\bar{\phi}(t)$, the implication is that so is each of the
trajectories in the ensemble. The generic classical trajectory is
the one whose value is equal to the RMS value of the distribution
at large times as given by Eq.~(\ref{27}). Now by setting $\Delta
\tau = \Delta \phi / \dot \phi$, we find
\beq \label{hdt}
H \Delta \tau= {2 \sqrt{\xi} \, \Delta \phi \over H \mu_\phi}
\left( {9 \pi^3 \over e} \right)^{1/4}
\left[
{\exp\left[{1 \over 4 \xi^2}\left(e^{-\sqrt{2}\xi}+\sqrt{2}\xi
-e^{-4 \xi N_{fl}/3w}- {4 \xi N_{fl}\over 3 w}\right)\right]\over
(1-e^{-4 \xi N_e/3w})}\right]
\eeq
Here $N_{fl}$ denotes the time that the initial condition is
established, as the wave goes outside the Hubble length during
inflation, and $N_e$ denotes the time at which inflation ends;
both times are measured in units of $e$-foldings. This formula
applies for the first type of models; the exponent has a $2
\mu_\psi^2 N/3$ in the second type.

The effect of a fluctuation is to determine the time and the
value of $\phi$ at which diffusion ends and the classical
evolution takes over ($N_e$). The direct change in the time at
which classical evolution begins translates into a difference in
time at which inflation ends.  We expand the above answer for
small $\xi$ to get
\beq
H \Delta \tau=(9\pi^3)^{1/4}{3 \over 2N_e}{\Delta \phi \over H}
w^{3/2}e^{-2N_{fl}^2/9w^2} \label{40}
\eeq
where $N_{fl}$ is the time at which the fluctuation occurs.  This
falls off from the peak at $N_{fl}=0$ like a Gaussian with width
$\sqrt{9w^2/2}$.  What this tells us is that fluctuations formed
sufficiently early (or late) will not delay the onset of the
classical regime significantly and not give a significant
contribution to density fluctuations.  However, fluctuations
formed during the diffusive growth regime near $N=0$ will create
far too large density perturbations.  These fluctuations must be
such that they are not relevant to observable scales. We can
observe back to about 40 $e$-folds before the end of inflation,
so we require that the fluctuations formed at this time were
sufficiently small, or that inflation must end by 40 $e$-folds
beyond the time when the $\phi$ fluctuations satisfy the
experimental bound.  From Eqn. \ref{40}, one can deduce this time
is approximately $N_{df}=-5w$. It might be thought that another
solution is that inflation ends very slowly, so that 40 $e$-folds
before the end one is in the classical regime.  However, when
$\phi$ is evolving according to the classical equations of
motion, the scale dependence of density perturbations is much too
large.

So the number of $e$-folds beyond the time when density
fluctuations in $\phi$ are sufficiently small is
\beq
N=|N_{df}|+\sqrt{9 \over 8}w+N_1+N_2+N_3
\eeq
Density fluctuations on the scale of 1 Mpc are formed
\beq
N_{Mpc}=38+{2 \over 3}\log(M/10^{11}{\rm GeV})+{1 \over
3}\log(T_{RH}/10^7{\rm GeV})
\eeq
$e$-folds before the end of inflation. We require that $N$ is
less than $N_{Mpc}$.  By following through the above
calculations, one can see this gives the approximate constraint
$w\approx 1$.  The detailed application of the constraint gives
the constraints illustrated in Figures 1-3. These plots were made
assuming the larger decay rate. The total number of $e$-folds for
the same parameters is generally about 5 larger with $\Gamma_l$
which can be accomodated with a modest change in $\mu_\phi$.

\section{Black Holes?}

Because of the large peak in the density perturbation spectrum on
small length scales arising from the $\phi$ contribution, there
is a danger too many small black holes being created.  There are
fairly strong constraints on the fractional mass density in black
holes on small scales \cite{cgl}. We investigate these
constraints on our model in this section.

First we summarize the constraints.  In the paper of Carr,
Gilbert, and Lidsey, the constraints are presented in several
forms; one constraint is on the parameter $\delta$ which is
related to $\delta \rho/\rho$. In the mass range above $10^{30}$
gm there are bounds from CMB distortions constraining $\delta$ to
be less than about $10^{-2}$. In the mass range between $10^{30}$
gm and $10^{10}$ gm the bound on $\delta$ is approximately
$10^{-1.5}$. In \cite{cgl} a bound due to relics is deduced
constraining $\delta$ between about $10^{-2}$ and $10^{-1.5}$ in
this mass range. This constraint from relics is perhaps more
speculative than robust bounds from not exceeding critical
density, or that decaying black holes do not produce too much
entropy.

In order to apply these bounds, one needs to know the probability
of black hole creation as a function of $\delta$.  Based on
\cite{carr}, the bound is obtained from applying the formula for
the probability of a region of mass $M$ forming a primordial
black hole
\begin{equation} \label{bheq}
\beta_0(M)\approx \delta(M) \exp (-{\gamma^2 \over 2 \delta^2})
\end{equation}
where the equation of state when the perturbation enters
the horizon is $p=\gamma \rho$. For the scales which
are of interest to us, $\gamma=1/3$, which we will assume
in the equations below.

The above bound comes from considering a spherically symmetric
overdense region. The requirement is made that when the overdense
region stops expanding, it's size $S_c$ exceeds the Jeans radius
$R_J$ at this time $t_c$, in order to collapse against the
pressure.  To derive exact numerical bounds on $\delta$ requires
that this be the precise condition. Without solving the full
problem explicitly including the pressure effects near the
boundary it is difficult to state precisely the conclusion, which
gives rise to some overall uncertainty in the $\delta$ bound.
However, one should be able to obtain a conservative bound on
$\delta$ through this approximation. However, even using this
approximation, we find numerical discrepancies with the precise
production rate which would be predicted.  First, the relation
between $t_c$ and $t_0$ (the time at which the perturbation
begins to evolve separately from the homogeneous background in
which it is embedded) should be $t_c=2 t_0 \delta^{-1}$ (the 2
being omitted in Ref. \cite{carr}) and the relation between $R_J$
and $t_c$ should be $R_J=2 v_s \pi \sqrt{2/3} t_c$ rather than
$v_s 2 t_c$ (in actual fact the 2 was omitted but cancels later
on), where $v_s$ is the sound velocity.  Overall this translates
into the bound $\delta_0>v_s^2 2 \pi \sqrt{2/3} \left({m \over
m_0}\right)^{-2/3}$ (where the correct relation $S_c=R_0
\delta_0^{-1/2}$ has been substituted and the ratio $t_0/R_0$ has
been replaced by the appropriate mass ratio). The implication is
that the factor $\gamma^2 /2 $ in Eqn. \ref{bheq} should be
replaced by $4 \pi^2 \gamma^2/3$, which in turn decreases the
strength of the bound on $\delta$ by a factor of about 5. This
will of course also weaken the bound on the scalar index $n$
given in \cite{cgl}.

 Our spectrum is not scale invariant on these small length scales
which is important when calculating $\delta$ from $H \Delta \tau$.
However a very conservative upper bound on our spectrum is a
scale invariant spectrum starting at a small length scale (near
the peak of the Gaussian) and which is constant over smaller
wavelengths.  This spectrum would be a scale invariant spectrum
with a cutoff at large wavelength, and can readily be compared
with the analysis of Ref.~\cite{cgl}.  The normalization they
use for $\delta$ can be extracted from their Eqs.~(4.2)--(4.4),
which express the value of $\delta$ at the COBE scale in terms of
the underlying inflaton potential.  Assuming that $T/S \ll 1$,
their equations reduce to $\delta=0.99 V^{3/2}/V'$.  By comparing
with the relation $H \Delta \tau = H \delta \phi/\dot{\phi} = H^2
/ (2 \pi \dot \phi) $, one finds that $\delta$ can be related to
the fluctuations in $\tau$ by $\delta =.086 H \Delta \tau$.  We
can then apply formula \ref{hdt} to find that $\delta$
(normalized as above) never exceeds 0.02, in the parameter regime
presented in Figures 1-4.  Because the quoted constraint on
$\delta$ in \cite{cgl} appears be too strong by a factor of 5, we
conclude that we are consistent with reasonable estimates of the
black hole constraint. Therefore, even with a conservative
overestimate of $\delta$, the constraint from black holes is
satisfied. However, there can be a sizable fraction of matter in
black holes, which would be interesting to study in the future.

As a final comment, we remark that the bound from black holes is
somewhat weaker if inflation ends more quickly, which it does for
lower $w$. In fact, for $w \approx 0.5$ (corresponding to
$m_\phi$ about four times larger than assumed) inflation ends in
about 10 $e$-foldings. In this case, only the bounds from relics
would apply. As this bound is more speculative, it is possible
that even large perturbations on this scale would be acceptable.
In reality, smaller $w$, while decreasing the length of time for
inflation to end, also decreases the density perturbations.
Larger $m_\phi$ always leads to a smaller fraction of the
universe in black holes.

In summary, the black hole constraint is a serious constraint and
must be accounted for. This is another constraint which would
forbid large $w$, since the maximum value of $H \Delta \tau$
grows as $w^{1/2}$ (when the dependence of $N_e$ on $w$ is
accounted for).  However we have seen our model is safely within
the bounds given in the literature once we have applied the
$m_\phi$ bound given in Section 7.

However these bounds are not sufficiently precise at present and
it would be interesting to do a more accurate calculation of the
mass fraction in black holes both for our model and in
general.\footnote{We thank B. Carr for informing us that work is
in progress on this subject.}  The effects we have discussed
should weaken existing bounds.

\section{  Gravitino Constraint }

In this two field model of inflation, the source of entropy and
energy in the universe is the decay of the $\psi$ and $\phi$
fields.  The $\psi$ field decays first, as discussed earlier.  We
assume the decay products are quickly thermalized, giving an
effective temperature $T_\psi$. Sometime afterward the $\phi$
field reaches the minimum of its potential and begins to
oscillate about it.  We assume that these oscillations are damped
by gauge or Yukawa couplings.  As in Ref.~\cite{rt}, the decay
can occur through a coupling $g^2/\langle \phi \rangle \int d^4
\theta \chi^\dagger \chi \phi$ or through a direct Yukawa
coupling $\lambda \phi \phi_1 \phi_2$. This leads to a reheat
temperature equal to $max(g^{2/3} m_\phi^{5/6} M^{1/6},
m_\phi/\sqrt{\lambda})$, which is generally of order
$10^5$--$10^7$ \hbox{GeV}.  Since most of the energy of the
universe evolves from the coherent oscillations of the $\phi$
field, this reheat temperature sets the initial conditions for
the subsequent evolution.  As discussed in Ref.~\cite{moroi},
this  reheat temperature is low enough to avoid the
overproduction of gravitinos, even if gravitinos are as light as
100 GeV.

However, the initial temperature $T_\psi$ of the thermal plasma
of $\psi$ decay products can be as high as $10^{11}$ GeV, so the
production of gravitinos by this plasma must be examined.  In
this section we show that this constraint is never more
restrictive than the constraints already discussed.

Gravitinos are produced by scattering processes of the thermal
radiation, but interact at a rate suppressed by $m_{3/2}^2/M_p^2$.
They are potentially dangerous since they are not thermalized and
have a long lifetime.  The most stringent bounds are obtained by
considering the influence of these late decays on
nucleosynthesis.  The exact bound depends on the gravitino mass,
but for $100\hbox{ GeV} \lrtsim m_{3/2} \lrtsim 1\hbox{ TeV}$ it
is  $T_R<10^{7-9} { \rm GeV}$ \cite{moroi}.

Neglecting decay when considering gravitino production, one
writes the Boltzmann equation for the gravitino number density as
\cite{moroi}
\beq
{d n_{3/2} \over dt}+3Hn_{3/2}= \sigma_{\rm eff}\, n_{\rm rad}^2 \ ,
\eeq
where $n_{\rm rad}$ is the equilibrium number density of a single
species of scalar boson ($n_{\rm rad} =
\bigl(\zeta(3)/\pi^2\bigr) T^3$), and both the cross sections and
the multiplicity of species are accounted for by the factor
$\sigma_{\rm eff}$.  In terms of $Y_{3/2} \equiv n_{3/2}/n_{\rm
rad}$, we have
\beq \label{BoltzY}
{d Y_{3/2} \over dt}=\sigma_{\rm eff}\, n_{\rm rad} \ .\
\eeq
To a reasonable approximation $\sigma_{\rm eff}$ can be taken as
constant, although it does vary as the coupling constants run
and as species freeze out from the thermal equilibrium mix
\cite{moroi}.  For the standard case of a radiation-dominated
universe, $n_{\rm rad} \propto 1/R^3(t) \propto 1/t^{3/2}$ (where
$R(t)$ is the scale factor), so the total gravitino production
can be estimated by integrating Eq.~(\ref{BoltzY}) from the
initial reheat time $t_0$ to infinity.  This gives $Y_{3/2}
\approx 2 \sigma_{\rm eff}\, n_{{\rm rad},0}\, t_0 = \sigma_{\rm
eff} \, n_{{\rm rad},0} / H_0 \propto T_0$, where the subscript
$0$ refers to the time of reheating.  To obtain a reasonable
estimate of the present value of $Y_{3/2}$, one must divide this
value by a dilution factor to account for the production of
photons at times much later than $t_0$.  According to
Ref.~\cite{moroi}, the final result is
\beq \label{standard}
Y_{3/2}(T \ll 1 {\rm MeV})\approx 2.14 \times 10^{-11}
\left({T_0 \over 10^{10}\hbox{ GeV}}\right) \ .
\eeq

To derive a  conservative estimate for the gravitino
production of the $\psi$ decay products, we assume that the
energy released by the $\psi$ decay is approximately
equal to the energy stored in the oscillating
$\phi$ field when inflation ends.  According to the
numerical simulations of our model, the fraction of energy in the
$\psi$ field was generally less than this by a factor of at least $10^4$,
except in the case $M'=M_p$ and a slow decay rate,
in which case $\psi$ can store a substantial
 fraction of the energy at the end of inflation.    The universe then rapidly
become matter-dominated,
so $R(t) \propto t^{2/3}$.  Repeating the calculation for
$Y_{3/2}$ with this time evolution, one finds $Y_{3/2} = {2 \over
3} \, \sigma_{\rm eff} \, n_{{\rm rad},0} / H_0$, essentially the
same formula as above.

However, in this model there is an additional dilution of the
$\psi$ decay products, because the $\phi$ field behaves as a
coherent state of nonrelativistic particles for a time
$\Gamma_\phi^{-1}$, and then the $\phi$ particles decay to
produce radiation.  Before the $\phi$ particles decay, the energy
density $\rho_\psi$ of the $\psi$ decay particles (assumed to be
effectively massless) is suppressed relative to the
energy density $\rho_\phi$ of the $\phi$ field by one power of
the growth of the scale factor
between the times $t_0$ and $\Gamma_\phi^{-1}$, which is $1/(\Gamma_\phi
t_0)^{2/3}$.  When the
$\phi$ particles decay to radiation, the number of radiation
particles produced exceeds the number of $\psi$ decay particles by
$(\rho_\phi/\rho_\psi)^{3/4} \approx 1/(\Gamma_\phi t_0)^{1/2}$.
Relating $\Gamma_\phi$ to the $\phi$ reheat temperature $T_R
\approx 10^7$ GeV and
taking $t_0 \approx 1/m_\phi$, the dilution factor is found to be
approximately $(M_P m_\phi / T_R^2)^{1/2} \approx 10^4$.
Incorporating this extra dilution factor into
Eq.~(\ref{standard}), we find that gravitino production from
$\psi$ decay products give
\beq
Y_{3/2}(T\ll 1 \hbox{ MeV})\approx 10^{-15} \left({T_\psi\over
10^{10} \hbox{ GeV}}\right) \ .
\eeq
where we have explicitly incorporated our
assumption that  $\psi$ and $\phi$ initially carry comparable
energy.

Thus, a $\psi$ reheat temperature of $10^{11}$ GeV produces no
more gravitinos then a final reheat temperature of $10^7$
\hbox{GeV}.  Thus,   we
find that no further constraints need to be imposed.

\section{Baryogenesis}

In the context of late scale inflation, it is worthwhile to
investigate the question of how baryons are created. There are
essentially two known possibilities. Electroweak baryogenesis
\cite{ew} is possible since the reheat temperature will generally
be above the weak scale.  Alternatively, a model in the context
of supersymmetry invites investigation into the Affleck Dine
scenario \cite{ad}.

In the Affleck Dine baryogenesis scheme, a field $\chi$ which
carries baryon number acquires a large displacement relative to
its true minimum somewhere during the early evolution of the
universe. If the interactions which drive the field to the true
minimum are CP and $B$ violating, the field will store baryon
number, and subsequently decay to baryon number carrying
particles.

In our model, in principle, the fields $\phi$ or $\psi$ could be
the AD fields. However this does not work. The problem is that
fields which carry baryon number will generally also carry
charge, so that $\phi$ is not a good possibility since charge (or
color) would be spontaneously broken by the vacuum. Although
$\psi$ is in principle a candidate, the ratio of baryon number
stored by the $\psi$ field to entropy will be too small.

This can be deduced from a detailed study of the $\psi$ field.
The first point to observe is that the potentials we have studied
to now are $B$ and CP conserving.  This is because we have
neglected the soft ``$A$" type terms and possible cross terms
which can violate CP.  When these are included, we find there can
cause a small change in the detailed evolution of the $\phi$
field.  At
the time the $A$ (CP and $B$ violating terms) are large, the
$\psi$ field only carried a small fraction of the energy of the
universe.

However, a separate flat direction which plays the role of the AD
field would work.  If the AD field is independent of the
inflation fields, it can then have large expectation value
through the final stages of inflation.  If $H$ is somewhat larger
than $m_{AD}$, the analysis is similar to that in Ref. \cite{drt}
where a much larger $H$ was assumed.  The baryon to entropy ratio
is approximately
\beq
{n_b \over s}\approx {n_b \over n_{AD}}{T_R \over m_{AD}}
{\rho_{AD} \over T_R^4}
\eeq
where $n_b/n_{AD}$ gives the baryon to particle number ratio in
the AD field, and should be order unity if the potential for the
field is $B$ and CP violating.  The last factor $\rho_{AD}/T_R^4$
is determined by the amplitude of the AD field at the time it
evolves towards its true minimum, which is determined by higher
dimension operators in the potential \cite{drt}.  One can readily
obtain acceptable values for the baryon density if the dimension
of the operator in the superpotential which lifts the flat
direction is greater than 4. The lower reheat temperature
expected in these models requires a correspondingly larger factor
$\rho_{AD}/T_R^4$, so a dimension 4 operator in the
superpotential which lifts the AD field is insufficient.

\section{Discussion}

There are several comments to make about the models we have
considered. First there is the fact that we are considering very
late inflation. We do not address the question of why the
universe has lasted to this point \cite{in}.  We have only
addressed the question of a late inflationary epoch responsible
for solving the horizon and flatness problems and for generating
the necessary density fluctuations.

 One possible solution is that the initial $\psi$ value exceeds
$M_p$.  In this case, it is possible that chaotic inflation could
solve the problem raised above.  However, subsequent to this
stage of inflation, one would expect inflation as described in
this paper which would create density fluctuations of the right
size.

Another point we have not addressed is what our models look like
when embedded in supergravity theories. Our point of view
throughout this paper is to regard the theory as an effective
theory expanded in powers of $M_p$ (and $M'$). We have neglected terms which
are suppressed by higher powers of the Planck scale.  For the
same reason we have assumed a minimal Kahler potential when
deriving the potential. From this point of view, any secondary
minima which occur in supergravity for field values exceeding
$M_p$ are not to be trusted.

  One aspect of our models which is important is the requirement
of renormalizable couplings of the fields in the $\phi$ direction
in order to obtain sufficiently high reheat temperature. For this
reason we expect it is more likely that the fields $\phi$ and
$\psi$ correspond to flat directions of a renormalizable theory
(along the lines discussed in Section 6) than to true moduli
fields (of string theory) \cite{lance}.

In our models, we saw that there was usually some small but not
very small number.  Either $M'$ is $M_p$ in which case $\mu_\psi$
and $1/\mu_\phi$ are of order 100, or $M'$ is smaller than $M_p$
which permits $\mu_\psi$ and $\mu_\phi$ closer to unity. Another
possibility is that the small number is related to a Yukawa
coupling. We discuss each of these possibilities in turn.

It is well known that there can be $H$ dependent correction to
the soft supersymmetry breaking masses at early times when $H$
exceeds $m_{3/2}$ \cite{drt,largeh}.  This means that small
$\mu_\psi$ is necessarily obtained by tuning.  Stewart\cite{s1}
has presented criteria which are sufficient for the cancellation
of supergravity corrections to the inflaton mass, so that even a
large ratio is technically consistent.  However these conditions
will only work when the scale of inflation is above the
supersymmetry breaking scale. One therefore needs to invoke a new
mass scale. Generally Stewart chooses the scale of gaugino
condensation. It is hard to see how this scale is realized in an
actual model although it could present an interesting
alternative. There is a tradeoff between the complexity of the
model and the ``naturalness" of taking $m_\psi$ somewhat smaller
than $H$.

In the models where $\mu_\psi$ is small, we found that
consistency of the model required that $\mu_\phi$ is large, with
the product $\mu_\phi \mu_\psi$ being a number of order unity.
It might be thought that this large value of  $\mu_\phi$ could be
  explained as due to large $H$ dependent mass
corrections. However, it is not possible to introduce a large
$\mu_\phi$ without the tuning parameter appearing in some other
unnatural feature of the potential. For example, a large $H$
dependent mass could introduce a new minimum for $\phi$ which is
closer to the origin so that the VEV of $\phi$ is correspondingly
smaller than $M_p$.

 We have seen however that the tuning of mass ratios is
significantly reduced if we accept a higher scale than the
conventional intermediate scale as setting the overall energy
density.  However if this scale has anything to do with visible
supersymmetry breaking, the highest scale possible is probably
the gaugino condensate scale. The predictions for $n$ and $R$
would be very similar, so the general test for this class of
models would still be valid.

We regard the small tuning of parameters as a necessary aspect of
the models with $M'\approx M_p$. The necessary numbers may or may
not be present. The tuning is certainly much smaller than in a
typical inflation model.

On the other hand, $M'$ might be smaller. This requires the
presence of another mass scale in the theory.  If this lower
scale exists, one can obtain $\mu_\psi$ and $\mu_\phi$ closer to
unity.

The other possibility is that there is a small Yukawa coupling.
This is probably not such a bad possibility. First of all, the
necessary coupling is no smaller than known Yukawa couplings and
might even be related to them as in the model of Section 6 and
obvious variants.  Second, known Yukawas can be derived as the
ratio of mass scales. In the SU(6) model we discussed this has
been done in Ref. \cite{bbh,cr}.  It is not unreasonable to think
there might be such effective Yukawa couplings in hidden sectors
of the theory, as well as in the single known sector.

In most known hybrid inflation models other than the one we
discussed, the $\psi$ field is very light, while the $\phi$
field is very heavy, of order $M$. This is a more serious
technical problem since radiative corrections will generally give
$\psi$ too large a mass \cite{shafi}.  Even if the model is
supersymmetric, supersymmetry breaking during inflation would
induce a large mass for $\psi$.  Although at tree level
$Str(M^2)=0$, this is not sufficient to prevent radiative
corrections at higher loop order. One can perhaps allow for such
a hierarchy, but at the expense of additional complexity and mass
scales. A chief advantage of our model is that both $m_\phi$ and
$m_\psi$ are of order the soft supersymmetry breaking scale so
this problem does not arise.

We view our model as the simplest illustration that flat
directions of supersymmetric theories are consistent with the
requirements of inflation when one allows for more than one field
in the inflation sector.  It is likely that the small parameters
which might be required (of order 0.01 to 0.1) are present.
Alternatively there might be more subtle mechanisms at work.
Either way, one would conclude that the scale of inflation is
very low. Even allowing inflation to be determined by the higher
gaugino condensate scale, one would conclude that $H$ during
inflation is between $10^4$ and $10^7$ GeV, and tensor
perturbations are small.

It is important that there are observational consequences to this
type of model. The combination of measuring the scalar index $n$
and $T/S$ should either rule out or encourage belief in the
mechanism at work here.  As discussed in Section 3,
\beq
n=1-3\left({V' \over V}\right)^2+2{V''\over V}
\eeq
Because the second term is negligible in models of the sort we
are considering, where $\psi$ at the end of inflation is much
less than $M_{Pl}$, it is only the last term which causes the
deviation of $n$ from unity.  If the dependence on $\psi$ is
dominated by a mass term, as in the model of Section 3, the
correction to $n$ will be positive (but small). We then expect
$n$ greater than or equal to unity, and $T/S$ to be small.

\section{Conclusion}
We have shown that with more than one field it is possible to
construct models of inflation with no small parameters.
Furthermore, the mass scales which seem to most naturally appear
in these models are of order $m_{3/2}$, about 1 TeV, and $M_I$,
about $10^{11}$ GeV, leading to a natural association with
supersymmetric models.  These models give rise to the correctly
normalized density perturbations, even though the Hubble constant
is quite low, of order $10^4$ GeV, because the value of the
inflaton field at the end of inflation is much lower than the
Planck scale.  The key to producing more such models is a
sensitive dependence of the $\phi$ potential on the value of the
$\psi$ field, so that the motion of the $\psi$ field can trigger
the end of inflation while its value is small.

It seems that multifield models are probably the most natural
models which can implement inflation with weak scale Hubble
constant, and that furthermore, these are probably the most
natural inflation models in that they involve no new small
parameters. The requisite small parameters arise naturally from
the ratio of mass scales. These models have the further
advantages that they can be explicitly realized and one can
calculate the relevant parameters for any particular
implementation. They might even occur in simple extensions of the
MSSM.  

Perhaps the most important property of a model is its
testability, and our proposed models have several characteristics
that are in principle observable.  The scalar index $n$ which
characterizes the scale dependence of density perturbations is
always greater than unity.  It is very close to unity for the
model of Sec.~III with $M'$ at the Planck or GUT scale, but for
$M'$ at the intermediate scale or for the model of Sec.~IV, it
could be as large as 1.2 for the parameters
shown in our plots.  In all cases tensor perturbations are
negligible.  An especially distinctive feature is a large spike
in the density perturbation spectrum at present wavelengths of
about 1 Mpc or less.

\subsection*{Acknowledgements}
 We are very grateful to Andrew Liddle, David Lyth, and Ewan
Stewart for discussions and the Aspen Center for Physics where
these discussions took place.  We also thank Sean Carroll, Csaba
Cs\'aki, Arthur Kosowsky, Andrei Linde, and Bharat Ratra for
their comments.  We thank Bernard Carr, Jim Lidsey, Avi Loeb,
Paul Schechter,  and
Paul Steinhardt for discussions and correspondence about black
holes. We also thank Krishna Rajagopal for his comments on the
manuscript.

\end{document}